\documentclass[11pt,a4paper]{article}
\usepackage[hyperref]{acl2019}
\usepackage{times}  %Required
\usepackage{url}  %Required
\usepackage{booktabs} % For formal tables
\usepackage{subcaption}
\usepackage{graphicx}
\usepackage{amsmath}
\allowdisplaybreaks
\usepackage{amssymb}
\usepackage[inline]{enumitem}
\usepackage{makecell}
\usepackage{multirow}
\usepackage{color}
\usepackage{latexsym}

\aclfinalcopy % Uncomment this line for the final submission
\def\aclpaperid{762} %  Enter the acl Paper ID here

\title{Predicting TED Talk Ratings from Language and Prosody}

\author{Md Iftekhar Tanveer$^1$, Md Kamrul Hassan$^2$, Daniel Gildea$^3$, M. Ehsan Hoque$^4$ \\
    University of Rochester\\
    {\tt \{$^1$itanveer,$^2$mhasan8,$^3$gildea,$^4$mehoque\}@cs.rochester.edu}}

\date{}

\begin{document}
\maketitle
\begin{abstract}
We use the largest open repository of public speaking---TED Talks---to predict the ratings of the online viewers. Our dataset contains over 2200 TED Talk transcripts (includes over 200 thousand sentences), audio features and the associated meta information including about 5.5 Million ratings from spontaneous visitors of the website. We propose three neural network architectures and compare with statistical machine learning. Our experiments reveal that it is possible to predict all the 14 different ratings with an average AUC of 0.83 using the transcripts and prosody features only. The dataset and the complete source code is available for further analysis.
\end{abstract}

\section{Introduction}
Imagine you are a teacher, or a corporate employee, or an entrepreneur. Which soft skill do you think would be the most valuable in your daily life? According to an article in Forbes~\cite{Gallo2014a}, 70\% of employed Americans agree that public speaking skills are critical to their success at work. Yet, it is one of the most dreaded acts. Many people rate the fear of public speaking even higher than the fear of death~\cite{Wallechinsky2005}. As a result, several commercial products are being available nowadays to come up with automated tutoring systems for training public speaking. Predicting the viewer ratings is an essential component for the systems capable of tutoring oral presentations. 

We propose a framework to predict the viewer ratings of TED talks from the transcript and prosody component of the speech. We use a dataset of $2233$ public speaking videos accompanying over $5$ million viewer ratings. The viewers rate each talk on 14 different categories. These are---\emph{Beautiful}, \emph{Confusing}, \emph{Courageous}, \emph{Fascinating}, \emph{Funny}, \emph{Informative}, \emph{Ingenious}, \emph{Inspiring}, \emph{Jaw-Dropping}, \emph{Long-winded}, \emph{Obnoxious}, \emph{OK}, \emph{Persuasive}, and \emph{Unconvincing}. Besides, the complete manual transcriptions of the talks are available. As a result, this dataset provides high-quality multimedia contents with rich ground truth annotations from a significantly large number of spontaneous viewers. We release the data and the complete source code for future scientific exploration~\footnote{Link to source code blinded for author anonymity}.

TED talks are edited production videos. They contain numerous changes in the camera angles, clips from the presentation slides, reactions from the audience, etc. To avoid these extraneous features and to focus only on the speech, we remove the visual elements from the data. We use only the transcripts and the processed audio features (pitch, loudness etc.) in our experiments. However, the links to the original TED talks are preserved in the dataset. Therefore, it is possible to retrieve the visual elements if necessary.

We utilize three neural network architectures in our experiments. Our results show that the proposed solutions always outperform (AUC $0.83$) the baseline approaches (AUC $0.78$) for predicting the TED talk ratings.

\section{Background Research}
An example of behavioral prediction research is to \emph{automatically grade essays}, which has a long history~\cite{valenti2003overview}. Recently, the use of deep neural network based solutions~\cite{alikaniotis2016automatic,taghipour2016neural} are becoming popular in this field. \citet{farag2018neural} proposed an adversarial approach for their task. \citet{jin2018tdnn} proposed a two-stage deep neural network based solution. Predicting \emph{helpfulness}~\cite{martin2014prediction,yang2015semantic,liu2017using,chen2018cross} in the online reviews is another example of predicting human behavior. In general, behavioral prediction encompasses numerous areas such as predicting \emph{outcomes in job interviews}~\cite{Naim2016}, \emph{hirability}~\cite{Nguyen2016}, \emph{presentation performance}~\cite{Tanveer2015,Chen2017a,Tanveer2018} etc.

Research has been conducted on predicting various aspects of the TED talks. \citet{Chen2017} analyzed the TED Talks for humor detection. \citet{Liu2017} analyzed the transcripts of the TED talks to predict audience engagement in the form of applause. \citet{Haider2017} predicted user interest (engaging vs. non-engaging) from high-level visual features (e.g., camera angles) and audience applause. \citet{Pappas:2013:SAU:2484028.2484116} proposed a sentiment-aware nearest neighbor model for a multimedia recommendation over the TED talks. \citet{bertero2016long} proposed a combination of Convolutional Neural Network (CNN) and Long-short Term Memory (LSTM) based framework to predict humor in the dialogues. \citet{jaech2016phonological} analyzed the detection performance of phonological puns using various natural language processing techniques. \citet{weninger2013words} predicted the TED talk ratings from the linguistic features of the transcripts. This work is similar to ours. However, they did not use neural networks and thus obtained similar performance to our baseline methods.

\begin{figure}
\centering
\includegraphics[width=1\linewidth]{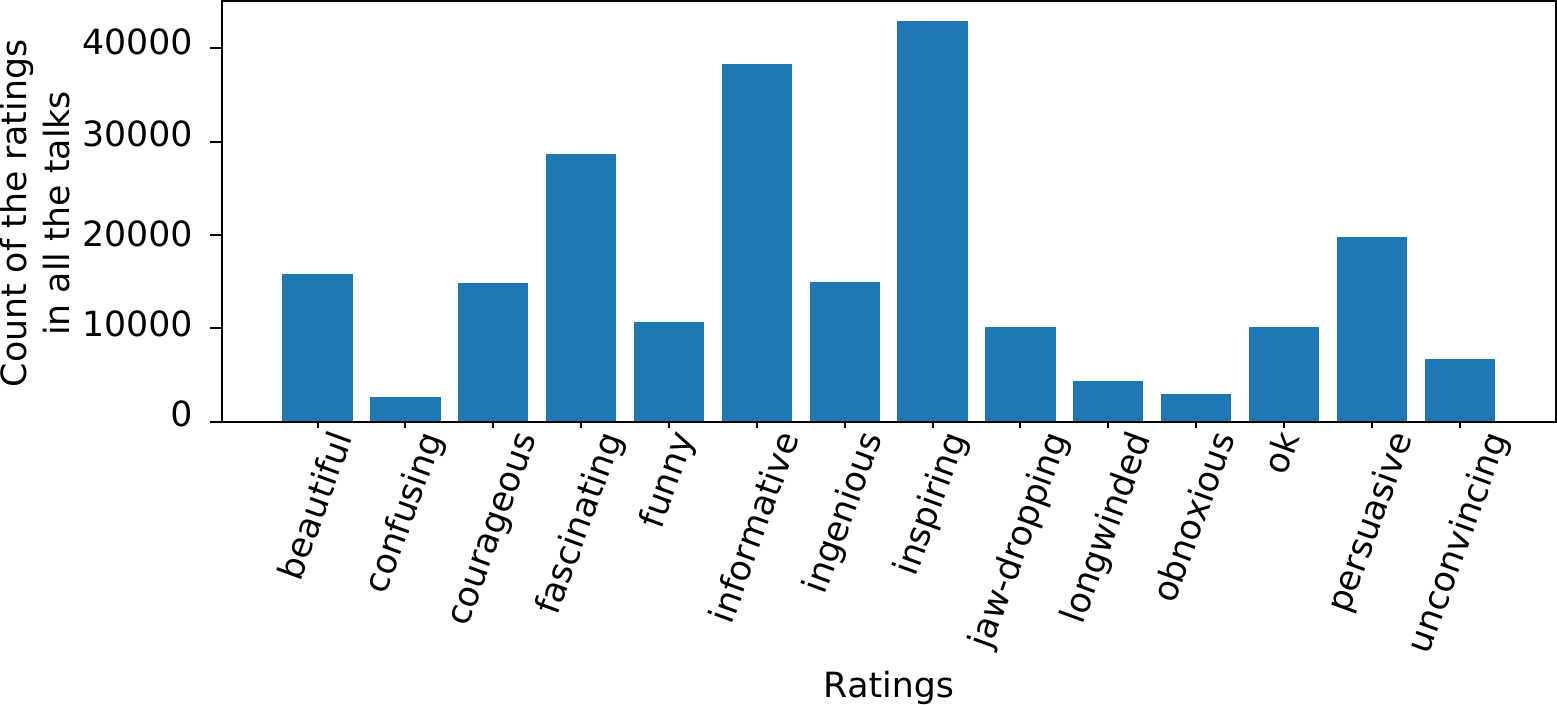}
\caption{Counts of all the 14 different rating categories (labels) in the dataset}
\label{fig:rating_counts}
\end{figure}
\section{Dataset}\label{sec:DatContents}
The data for this study was gathered from the \url{ted.com} website on November 15, 2017. We removed the talks published six months before the crawling date to make sure each talk has enough ratings for a robust analysis. More specifically, we filtered any talk that--- \begin{enumerate*} \item was published less than 6 months prior to the crawling date, \item contained any of the following keywords: live music, dance, music, performance, entertainment, or, \item contained less than 450 words in the transcript.\end{enumerate*} This left a total of 2231 talks in the dataset.

We collect the manual transcriptions and the total view counts for each video. We also collect the ``ratings'' which is the counts of the viewer-annotated labels. The viewers can annotate a talk from a selection of 14 different labels provided in the website. The labels are not mutually exclusive. Viewers can choose at most 3 labels for each talk. If only one label is chosen, it is counted 3 times. We count the total number of annotations under each label as shown in Figure~\ref{fig:rating_counts}. The ratings are treated as the ground truth about the audience perception. A summary of the dataset characteristics is shown in Table~\ref{tab:datasize}.

\begin{table}
  \centering
  \begin{tabular}{ll}
    \toprule
    \textbf{Property}& \textbf{Quantity}\\
    \midrule
\textbf{Number of talks} & 2,231\\
\textbf{Total length of all talks} & 513.49 Hours\\
\textbf{Total number of ratings} & 5,574,444\\
\textbf{Minimum number of ratings} & 88\\
\textbf{Average ratings per talk} & 2498.6\\
\textbf{Total word count} & 5,489,628\\
\textbf{Total sentence count} & 295,338\\
  \bottomrule
\end{tabular}
  \caption{Dataset Properties}
  \label{tab:datasize}
\end{table}

The longer a TED talk remains in the web, the more views it gets. Large number of views also result in a large number of annotations. As a result, older TED talks contain more annotations per rating category. However, an old speech does not necessarily imply better quality. We normalize the rating counts of each individual talk as in the following equation:
\begin{equation}\label{eq:scaled_score}
r_{i,\text{scaled}} = \frac{r_i}{\sum_i{r_i}} 
\end{equation}
Where $r_i$ represents the count of the $i^{\text{th}}$ label in a talk. Let us assume that in a talk, $f_i$ fractions of the total viewers annotate for the rating category $i$. Then the scaled rating, $r_{i,\text{scaled}}$ becomes $\frac{f_iV}{\sum_i{f_iV}}=\frac{Vf_i}{V\sum_i{f_i}}$. This process removes the effect of \emph{Total Views}, $V$ as evident in Table~\ref{tab:corrcoef}. Scaling the rating counts removes the effects of \emph{Total Views} by reducing the average correlation from $0.56$ to $-0.03$. This also removes the effect of the \emph{Age of the Talks} by reducing the average correlation from $0.15$ to $0.06$. Therefore, removing $V$ reduces the effect of the \emph{Age of the Talks} in the ratings.
\begin{table}
\centering
\begin{tabular}{lrrrr} 
\toprule
                       & \multicolumn{2}{l}{\textbf{\makecell{Total Views}}} & \multicolumn{2}{l}{\textbf{\makecell{Age of Talks}}}  \\
                       & \textbf{noscale} & \textbf{scale}        & \textbf{noscale} & \textbf{scale}          \\ 
\midrule
\textbf{Beaut.}     & 0.52             & 0.01                  & 0.03             & -0.14                   \\
\textbf{Conf.}     & 0.39             & -0.12                 & 0.27             & 0.20                    \\
\textbf{Cour.}    & 0.52             & -0.003                & 0.01             & 0.15                    \\
\textbf{Fasc.}   & 0.78             & 0.05                  & 0.15             & 0.06                    \\
\textbf{Funny}         & 0.57             & 0.14                  & 0.10             & 0.10                    \\
\textbf{Info.}   & 0.76             & -0.08                 & 0.07             & -0.19                   \\
\textbf{Ingen.}      & 0.59             & -0.06                 & 0.18             & 0.10                    \\
\textbf{Insp.}     & 0.79             & 0.1                   & 0.05             & -0.15                   \\
\textbf{Jaw-Dr.}  & 0.51             & 0.1                   & 0.18             & 0.23                    \\
\textbf{Long.}    & 0.44             & -0.17                 & 0.36             & 0.31                    \\
\textbf{Obnox.}     & 0.27             & -0.11                 & 0.19             & 0.17                    \\
\textbf{OK}            & 0.72             & -0.16                 & 0.21             & 0.14                    \\
\textbf{Pers.}    & 0.72             & -0.01                 & 0.12             & 0.02                    \\
\textbf{Unconv.}  & 0.29             & -0.14                 & 0.18             & 0.15                    \\ 
\midrule
\textbf{Avg.}       & 0.56             & -0.03                 & 0.15             & 0.06                    \\
\bottomrule
\end{tabular}
\caption{Correlation coefficients of each category of the ratings with the \emph{Total Views} and the \emph{``Age'' of Talks}}
\label{tab:corrcoef}
\end{table}

In our experiments, we scale and binarize the rating counts by thresholding over the median value which results in a $0$ and $1$ class for each category of the ratings. The dataset contains the complete original information as well as the scaled and binarized versions of the ratings.

\section{Network Architectures}
We implemented three neural networks for comparison of their performance with the statistical machine learning techniques in predicting the viewer ratings. The architectures of these models are described in the following subsections. All these models are multi-label binary classifiers designed to capture sentence-wise patterns in the TED talks that contribute to the prediction of the rating labels.

\subsection{Word Sequence Model}
\begin{figure}
\centering
\includegraphics[width=1\linewidth]{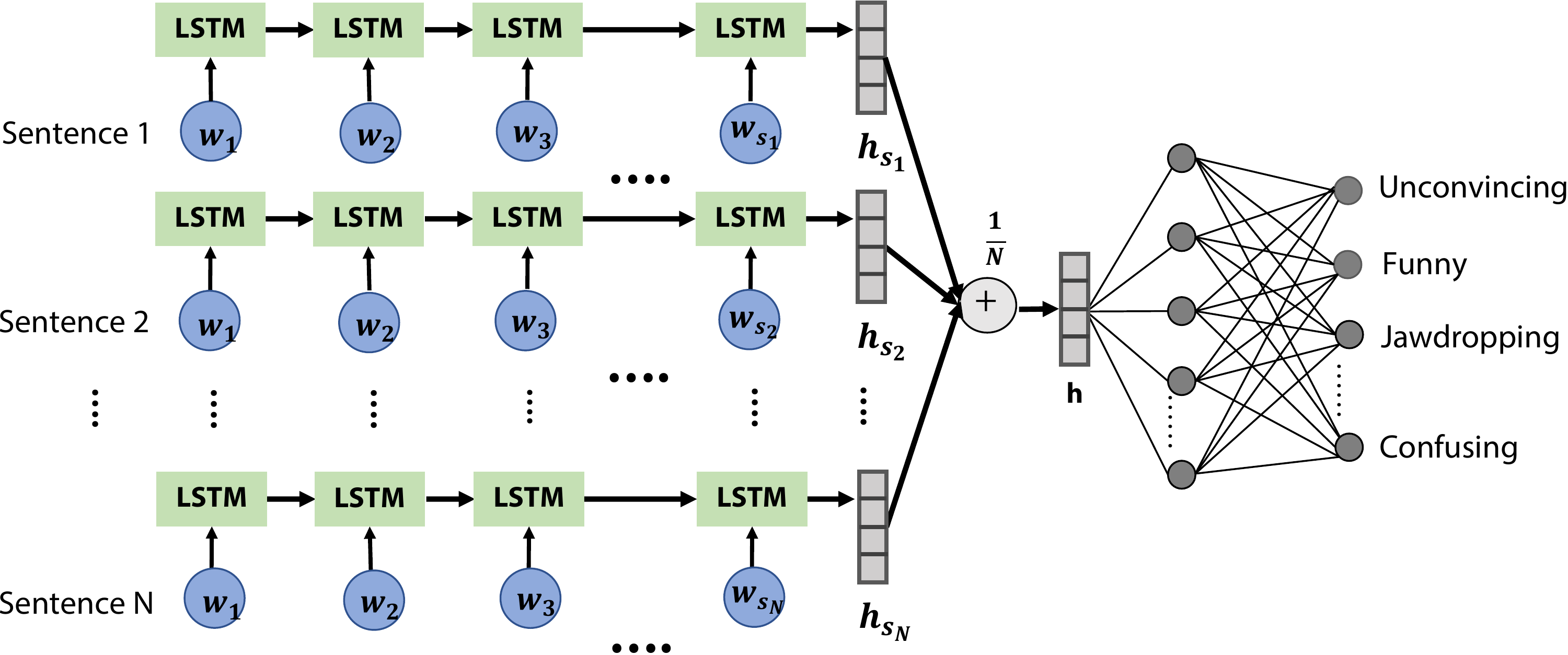}
\caption{An illustration of the Word Sequence Model}
\label{fig:model_word_seq}
\end{figure}

A pictorial illustration of this model is shown in Figure~\ref{fig:model_word_seq}. Each sentence, $s_j$ in the transcript is represented by a sequence of words-vectors\footnote{In this paper, we represent the column vectors as lowercase boldface letters; matrices or higher dimensional tensors as uppercase boldface letters and scalars as lowercase regular letters. We use a prime symbol ($'$) to represent the transpose operation.}, $\mathbf{w}_1,\mathbf{w}_2,\mathbf{w}_3,\dots,\mathbf{w}_{n_j}$. Here, each $\mathbf{w}$ represents the pre-trained, 300-dimensional GLOVE word vectors~\cite{pennington2014glove} corresponding to the words in the sentence. We use a Long-Short-Term-Memory (LSTM)~\cite{Hochreiter1997} neural network to obtain an embedding vector, $\mathbf{h}_{s_j}$, for the $j^{\text{th}}$ sentence in the talk transcript. These vectors ($\mathbf{h}_{s_j}$) are averaged and passed through a feed-forward network to produce a 14-dimensional output vector corresponding to each category of the ratings. An element-wise sigmoid ($\sigma(x) = \frac{1}{1+e^{-x}}$) activation function is applied to the output vector. The mathematical description of the model can be given using the following equations:
\begin{align}
&\mathbf{h}_{s_j} = \text{LSTM}(\mathbf{w}_1,\mathbf{w}_2,\mathbf{w}_3,\dots,\mathbf{w}_{n_j})\\
&\mathbf{h} = \frac{1}{N}\sum_{j=1}^N\mathbf{h}_{s_j}\\
&\mathbf{r} = \sigma(\mathbf{W}\mathbf{h} + \mathbf{b}_r)
\end{align}
Here, $\mathbf{h}_{s_j}$ represents the the last recurrent state for the sentence $j$. $N$ represents the total number of the sentences in the transcript. We use zero vectors to initialize the memory cell ($\mathbf{c}_0$) and the hidden state ($\mathbf{h}_0$).

\subsection{Dependency Tree-based Model}\label{sec:deptree_model}
\begin{figure}
\centering
\includegraphics[width=1\linewidth]{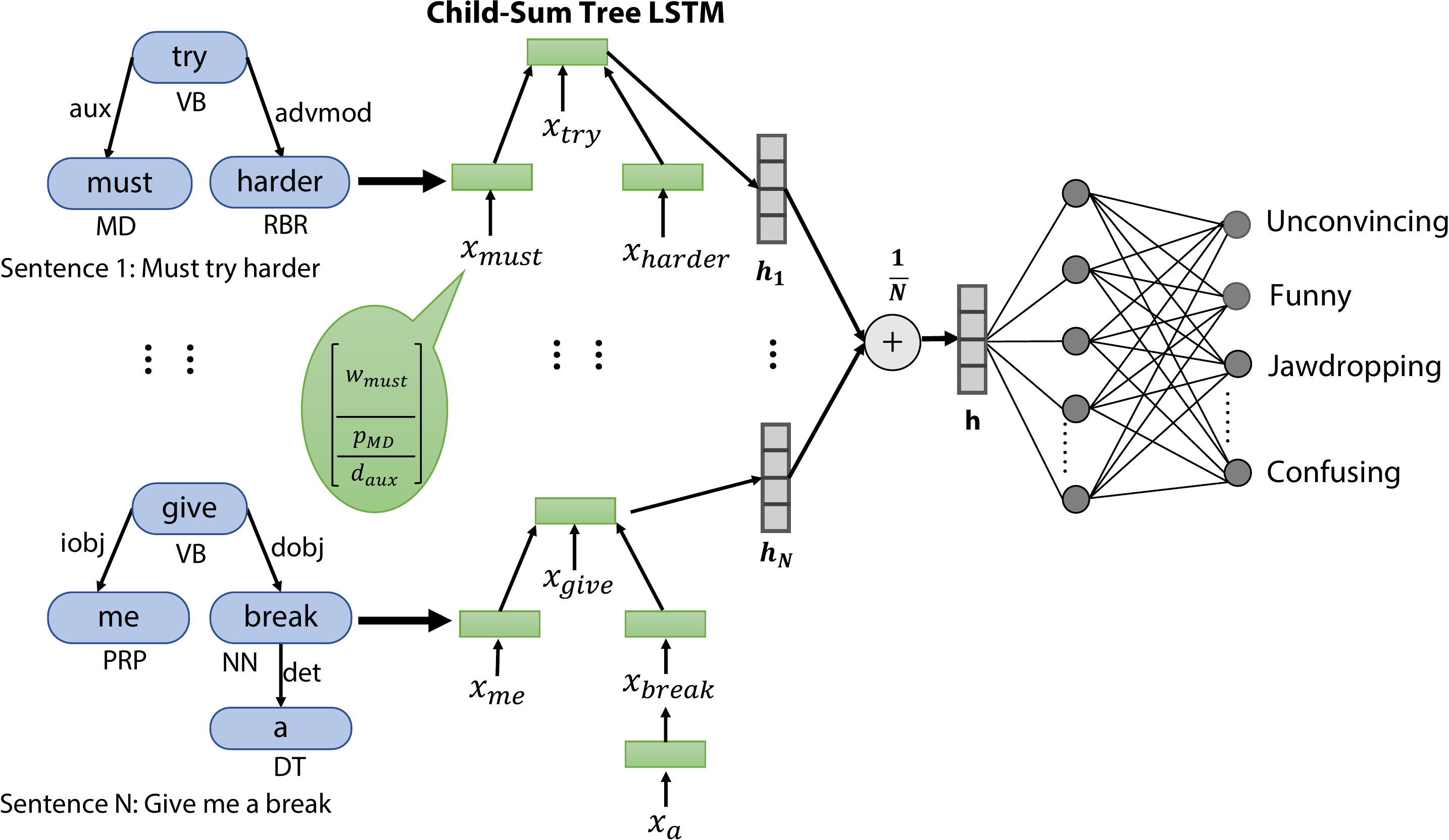}
\caption{An illustration of the Dependency Tree-based Model}
\label{fig:model_dep_tree}
\end{figure}

We are interested to represent the sentences as hierarchical trees of dependent words. We use a freely available dependency parser named SyntaxNet\footnote{https://opensource.google.com/projects/syntaxnet}~\cite{Andor2016} to extract the dependency tree corresponding to each sentence. The child-sum TreeLSTM~\cite{Tai2015} is used to process the dependency trees. As shown in Figure~\ref{fig:model_dep_tree}, the parts-of-speech and dependency types of the words are used in addition to the GLOVE word vectors. We concatenate a parts-of-speech embedding ($\mathbf{p}_i$) and a dependency type embedding ($\mathbf{d}_i$) with the word vectors. These embeddings are learned through back-propagation along with other free parameters of the network. The complete mathematical description of this model is as follows:
\begin{align}
&\mathbf{x'}_t = [\mathbf{w}'_t, \mathbf{p}'_t, \mathbf{d}'_t]\label{eq:concat}\\
&\mathbf{\tilde{h}}_t = \sum_{k\in C(t)}\mathbf{h}_k\label{eq:treelstm_first}\\
&\mathbf{i}_t =\sigma(\mathbf{U}_i\mathbf{x}_t+\mathbf{V}_i\mathbf{\tilde{h}}_{t} + \mathbf{b}_i)\label{eq:Vstart}\\
&\mathbf{f}_{tk}=\sigma(\mathbf{U}_f\mathbf{x}_t+\mathbf{V}_f\mathbf{h}_k + \mathbf{b}_f)\\
&\mathbf{u}_t =\tanh(\mathbf{U}_u\mathbf{x}_t+\mathbf{V}_u\mathbf{\tilde{h}}_{t} + \mathbf{b}_u)\\
&\mathbf{o}_t =\sigma(\mathbf{U}_o\mathbf{x}_t+\mathbf{V}_o\mathbf{\tilde{h}}_t + \mathbf{b}_o)\label{eq:Vend}\\
&\mathbf{c}_t =\mathbf{f}_{tk}\odot \mathbf{c}_k + \mathbf{i}_t\odot\mathbf{u}_t\\
&\mathbf{h}_t = \mathbf{o}_t\odot\tanh(\mathbf{c}_t)\label{eq:treelstm_last} \\
&\mathbf{h}_{s_j} = \mathbf{h}_{ROOT}\\
&\mathbf{h} = \frac{1}{N}\sum_{j=1}^N\mathbf{h}_{s_j}\\
&\mathbf{r} = \sigma(\mathbf{W}\mathbf{h} + \mathbf{b}_r)
\end{align}
Here, equation~\eqref{eq:concat} refers to the fact that the input to the treeLSTM nodes are constructed by concatenating the pre-trained GLOVE word-vectors with the embeddings of the parts of speech and the dependency type of a specific word. $C(t)$ represents the set of all the children of node $t$. The parent-child relation of the treeLSTM nodes come from the dependency tree. Notably, the memory cell and hidden states flow hierarchically from the children to the parent. Each node contains a forget gate ($\mathbf{f}$) for each child. Zero vectors are used as the children of the leaf nodes and the sentence embedding vector is obtained from the root node.

\subsection{Capturing the Patterns in Prosody}
We align the TED talk audio with its corresponding transcripts using forced alignment method~\footnote{https://github.com/JoFrhwld/FAVE/wiki/FAVE-align}. PRAAT~\footnote{http://www.fon.hum.uva.nl/praat/} is used to extract the pitch, loudness, and first three formants (frequency and bandwidth) sampled at a rate of 10Hz. We normalize these signals by subtracting the mean and dividing by the standard deviation over the whole video. These signals are then sentence-wise cropped based on the alignment data. We pad all the sentence-wise signal-clips to a length equal to the longest sentence in the transcript. This process constructs a signal of length $M$; where $M$ is the number of samples in the signal corresponding to the longest sentence. Each sample in the signal is an $8$ dimensional vector.

We use one dimensional Convolutional Neural Network (CNN)~\cite{LeCun2015} to extract the patterns within the pitch, loudness and formant as follows:
\begin{align*}
\mathbf{S}_{\text{out}}[f_{o},m] = &\sum_{f_i=1}^{F_{\text{in}}}\sum_{k=1}^K \mathbf{W}_F[f_o,f_i,k] \mathbf{S}_{\text{in}}[f_i,m-k]\\& + \mathbf{b}[f_o] \\ &\forall f_{o}\in{1,2, ..., F_{\text{out}}}\\
&\forall m\in{1,2, ..., M}
\end{align*}
Here $\mathbf{S_{\text{in}}}$ is the input signal, $\mathbf{S_{\text{out}}}$ is the output signal, $\mathbf{W}_F$ is the filter weights, $K$ is the receptive fields of the filters, $F_{\text{in}}$ is the dimension of the input signal, $F_{\text{out}}$ is the number of filters and $M$ is the signal length. $\mathbf{b}$ is a bias term. Both $\mathbf{W}_F$ and $\mathbf{b}$ are learned in training time through back-propagation. 

We use one dimensional Convolutional Neural Network (CNN)~\cite{LeCun2015} to extract the patterns within the \emph{prosody signal}---i.e. pitch, loudness, and the first three formants computed over small segments of the audio. The network consists of four 1D convolutional layers, each having a receptive field of 3. We use element-wise RELU ($\mathcal{R}(x) = \max(0,x)$) activation function to the output of each convolution layer. The lowest (closest to the input signal) two layers consist of 16 filters, and the upper two layers have 32 and 64 filters respectively. The second and third convolution layers are followed by max-pool layers of window size 2. The final convolution layer is followed by a max-pool layer having the window size equal to the length of the signal. Thus, the CNN outputs a 64-dimensional vector. This vector is concatenated with the sentence embedding vector obtained from the dependency tree-based model discussed in section~\ref{sec:deptree_model}. The concatenated vector is passed through two layers of fully connected networks to produce the probabilities of the ratings.

\section{Training the Networks}
We implemented the networks in pyTorch~\footnote{pytorch.org}. Details of the training procedure are described in the following subsections.

\subsection{Optimization}\label{sec:optimization}
We use multi-label Binary Cross-Entropy loss as defined below for the backpropagation of the gradients:
\begin{equation}
\ell(\mathbf{r},\mathbf{y}) = -\frac{1}{n}\sum_{i=1}^{n}(y_i\log(r_i) + (1-y_i)\log(1-r_i))
\end{equation}
Here $\mathbf{r}$ is the model output and $\mathbf{y}$ is the ground truth label obtained from data. $r_i$ and $y_i$ represent the $i^{\text{th}}$ element of $\mathbf{r}$ and $\mathbf{y}$. $n=14$ represents the number of the rating categories.

We randomly split the training dataset into 9:1 ratio and name them training and development subsets respectively. The networks are trained over the training subset. We use the loss in the development subset to tune the hyper-parameters, to adjust the learning rate and regularization strength, and to select the best model for final evaluation, etc. The training loop is terminated when the loss over the development subset saturates. The model parameters are saved only when the loss over the development subset is lower than any previous iteration.

We experiment with two optimization algorithms: Adam~\cite{Kingma2014} and Adagrad~\cite{Duchi2011}. The learning rate is varied in an exponential range from $0.0001$ to $1$. The optimization algorithms are evaluated with mini-batches of size $10$, $30$, and $50$. We obtain the best results using Adagrad with learning rate $0.01$ and in Adam with a learning rate of $0.00066$. The training loop ran for $50$ iterations which mostly saturates the development set loss. We conducted around $100$ experiments with various parameters. Experiments usually take about 48 hours to make 50 iterations over the dataset when running in an Nvidia K20 GPU.
\begin{figure}
\centering
\includegraphics[width=0.7\linewidth]{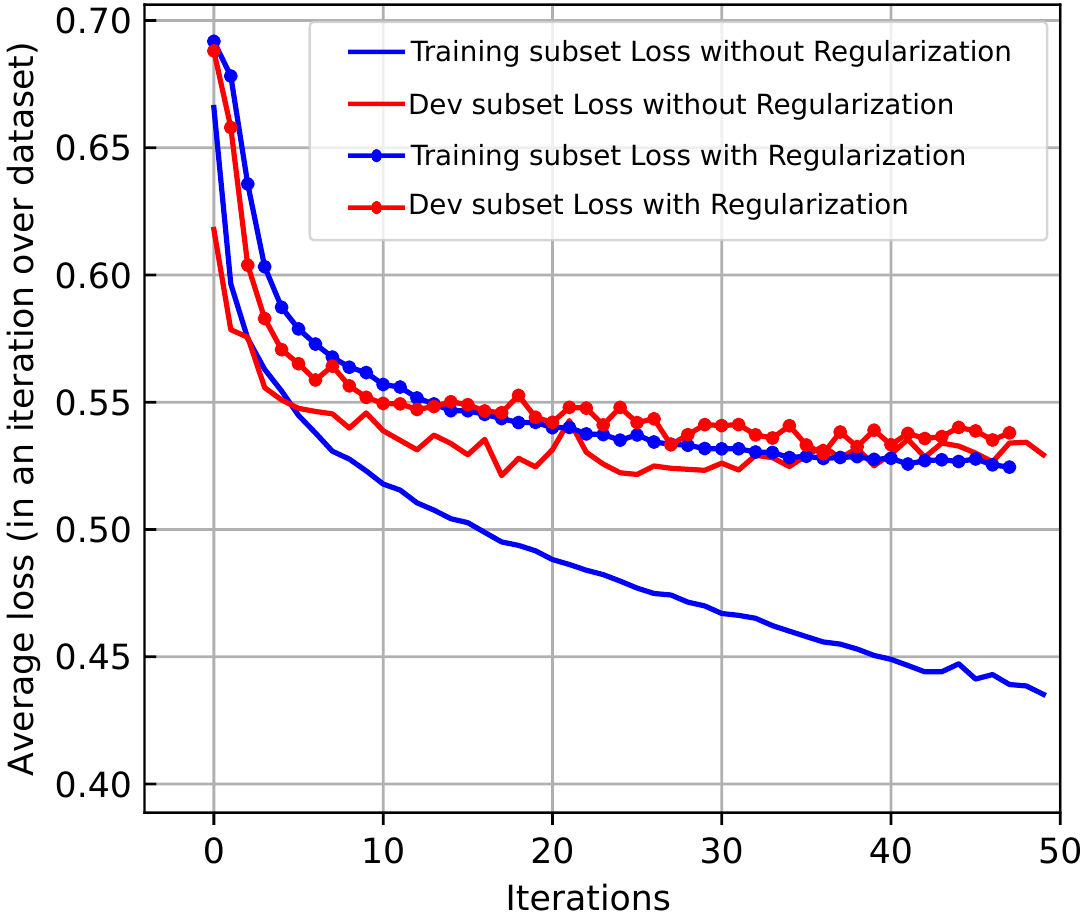}
\caption{Effect of Weight-Drop regularization on the training and development subset loss}
\label{fig:weight-drop}
\end{figure}
\subsection{Regularization}
Neural networks are often regularized using Dropout~\cite{Hinton2012} to prevent overfitting---where the elements of a layer's output are set to zero with a probability $p$ during the training time. A naive application of dropout to LSTM's hidden state disrupts its ability to retain long-term memory. We resolve this issue using the weight-dropping technique proposed by \citet{Merity2017}. In this technique, instead of applying the dropout operation between time-steps, it is applied to the hidden-to-hidden weight matrices~\cite{Wan2013}. The dropout probability, $p$ is set to $0.2$. Effect of the regularization is shown in Figure~\ref{fig:weight-drop}.

\section{Baseline Methods}
We compare the performance of the neural network models against several popular statistical classifiers.

\subsection{Feature Extraction}\label{sec:feat_extraction}
We use language, prosody, and narrative trajectory features that are used in similar tasks in the relevant literature.

\subsubsection{Language Features}
We use a psycholinguistic lexicon named ``Linguist Inquiry Word Count'' (LIWC)~\cite{Pennebaker-liwc01} for extracting language features. We count the total number of words under the 64 word categories provided in the LIWC lexicon and normalize these counts by the total number of words in the transcript. The LIWC categories include words describing function word categories (e.g., articles, quantifiers, pronouns), various content categories (e.g., anxiety, insight), positive emotions (e.g., happy, kind), negative emotions (e.g., sad, angry), etc. These features have been used in several related works~\cite{Ranganath2009,Zechner2009,Naim2016,Liu2017}.

\subsubsection{Prosodic Features}\label{subsec:audio_feat_extraction}
We extract several summary statistics from the pitch, loudness, and the first three formants extracted from the audio. These statistics are min, max, mean, 25th percentile, median, 75th percentile, standard deviation, kurtosis, and skewness. Additionally, we collect pause duration, the percentage of unvoiced frames, jitter (irregularities in pitch), shimmer (irregularities in vocal intensity), and percentage of breaks in speech. These features are used in several related works as well~\cite{Soman2009,Naim2016}.

\subsubsection{Narrative Trajectory}\label{subsec:audio_feat_extraction}
Tanveer et al.~\shortcite{Tanveer2018} proposed a set of features that can capture the ``narrative trajectory'' of the TED Talks. These features are constructed by extracting sentence-wise emotion (anger, disgust, fear, joy, or sadness), language (analytical, confidence, and tentative) and personality (openness, conscientiousness, extraversion, emotional range, and agreeableness) scores from a standard machine learning toolbox and then interpolating the sentence-wise scores to a signal of fixed size (e.g., 100 samples). These signals form several interesting clusters that can capture patterns of storytelling. The summary statistics of these signals are found to be good predictors of the TED talk ratings as well. We use the min, max, mean, standard deviation, kurtosis, and skewness of these signals. We use IBM Tone Analyzer~\footnote{https://www.ibm.com/watson/services/tone-analyzer/} to extract the sentence-wise scores.

\subsection{Baseline Classifiers}
We use the Linear Support Vector Machine (SVM)~\cite{Vapnik1964} and LASSO~\cite{Tibshirani1996} as the baseline classifiers. In SVM, the following objective function is minimized:
\begin{equation}
\begin{aligned}
& \underset{\mathbf{w}, \xi_i, b}{\text{minimize}}
& & \frac{1}{2}  \|  \mathbf{w} \|  + C \sum_{i = 1}^N \xi_i\\
& \text{subject to}
& & y_i \left(\mathbf{w}' \mathbf{x}_i - b\right) \geq 1 - \xi_i,  \ \forall i \\
&&&  \xi_i, \geq 0,  \ \forall i \\
\end{aligned}
\end{equation}
Where $\mathbf{w}$ is the weight vector and $b$ the bias term. $\|\mathbf{w}\|$ refers to the $\ell2$ norm of the vector $\mathbf{w}$. In these equations, we assume that the ``higher than median'' and ``lower than median'' classes are represented by $1$ and $-1$ values respectively.

We adapt the original Lasso~\cite{Tibshirani1996} regression model for classification purposes. It is equivalent to Logistic regression with $\ell1$ norm regularization. It works by solving the following optimization problem:
\begin{equation}
\begin{aligned}
&\underset{\mathbf{w},b}{\text{minimize}}
\quad \| \mathbf{w} \|_1 + k\\
& k=C\sum_{i=1}^N \log\left(\exp\left(-y_i\left(\mathbf{w}' \mathbf{x}_i + b \right)\right)+1\right) \\
\end{aligned}
\end{equation}
where $C > 0$ is the inverse of the regularization strength, and  $\| \mathbf{w} \|_1 = \sum_{j=1}^d |w_j|$ is the $\ell1$ norm of $\mathbf{w}$. The $\ell1$ norm regularization is known to push the coefficients of the irrelevant features down to zero, thus reducing the predictor variance.

Finally, the Ridge regression is essentially same as logistic regression with $\ell2$ regularization. The objective function is as below:
\begin{equation}
\begin{aligned}
&\underset{\mathbf{w},b}{\text{minimize}}
\quad \frac{1}{2}\|  \mathbf{w} \|+k \\
& k=C\sum_{i=1}^N \log\left(\exp\left(-y_i\left(\mathbf{w}' \mathbf{x}_i + b \right)\right)+1\right)
\end{aligned}
\end{equation}

\begin{table}
\centering
\begin{tabular}{lrrrr} 
\toprule
\textbf{Model}      & \multicolumn{1}{l}{\textbf{\makecell[l]{Avg.\\ AUC}}} & \multicolumn{1}{l}{\textbf{\makecell[l]{Avg.\\ F-sc.}}} & \multicolumn{1}{l}{\textbf{\makecell[l]{Avg.\\ Prec.}}} & \multicolumn{1}{l}{\textbf{\makecell[l]{Avg.\\Recall}}}  \\ 
\midrule
\textbf{Word Seq}            & 0.83                                     & 0.76                                        & 0.76                                           & 0.76                                         \\
\textbf{D.Tree}           & 0.83                                     & 0.77                                        & 0.77                                           & 0.77                                         \\
\textbf{D.Tree+Pr.} & 0.83                                     & 0.72                                        & 0.75                                           & 0.73                                         \\ 
\midrule
\textbf{\makecell[l]{Dep. Tree\\ (Unscaled)}} & 0.76                 & 0.70                 & 0.68               & 0.68                \\ 
\midrule
\textbf{LinearSVM}           & 0.78                                     & 0.71                                        & 0.71                                           & 0.71                                         \\
\textbf{Ridge}               & 0.78                                     & 0.71                                        & 0.71                                           & 0.71                                         \\
\textbf{LASSO}               & 0.77                                     & 0.70                                        & 0.70                                           & 0.70                                         \\
\midrule
\textbf{Weninger}                      & -- & 0.71                   &     --          & -- \\
\bottomrule
\end{tabular}
\caption{Average of several prediction performance metrics over 14 different ratings of TED talks}
\label{tab:avg_metrics}
\end{table}
\section{Experimental Results}\label{sec:exp_res}
We allocated $150$ randomly sampled TED talks from the dataset as a reserved test subset. All the results shown in this section are computed over this test subset. We evaluate the models by computing the values of four performance metrics---Area Under the ROC Curve (AUC), Precision, Recall, and F-score for all the 14 categories of the ratings. We compute averages of these metrics over all the rating categories that are shown in Table~\ref{tab:avg_metrics}. 

The first three rows represent the average performances of the Word Sequence model, the Dependency Tree based model, and the Dependency Tree model combined with CNN respectively. It is evident from the table that the neural networks outperform the baseline models in all the four metrics. These models were trained and tested on the scaled rating counts ($R_\text{scaled}$). We also trained and tested the dependency tree model with the unscaled rating counts ($4^\text{rd}$ row in Table~\ref{tab:avg_metrics}). Notably, the networks perform worse for predicting the unscaled ratings. We believe this is due to the fact that unscaled ratings are biased with the amount of time the TED talks remain online. This mixture of additional information makes it difficult for the neural networks to predict the ratings from transcript and prosody only.

We are surprised that adding the prosody does not improve the prediction performance. We think it is because TED Talks are highly rehearsed public speeches. It is likely that the change of prosody in most of the talks are acted, and therefore, it does not carry much information in addition to the talk transcripts. We believe it is a global artifact of the TED talk dataset.

\begin{table}
\centering
\begin{tabular}{lccc} 
\toprule
 \textbf{Ratings}      & \textbf{\makecell[l]{Word\\Seq.}}        & \textbf{\makecell[l]{Dep.\\ Tree}}       & \textbf{\makecell[l]{Weninger\\et al. (SVM)}}  \\ 
\midrule
\textbf{Beautiful}     & 0.88                     & \textbf{0.91}            & 0.80                     \\
\textbf{Confusing}     & 0.70                     & \textbf{0.74}            & 0.56                     \\
\textbf{Courageous}    & 0.84                     & \textbf{0.89}            & 0.79                     \\
\textbf{Fascinating}   & 0.75                     & 0.76                     & \textbf{0.80}            \\
\textbf{Funny}         & \textbf{0.78}            & 0.77                     & 0.76                     \\
\textbf{Informative}   & 0.81                     & \textbf{0.83}            & 0.78                     \\
\textbf{Ingenious}     & 0.80                     & \textbf{0.81}            & 0.74                     \\
\textbf{Inspiring}     & 0.72                     & \textbf{0.77}            & 0.72                     \\
\textbf{Jaw-dropping}  & 0.68                     & \textbf{0.72}            & \textbf{0.72}            \\
\textbf{Longwinded}    & \textbf{0.73}            & 0.70                     & 0.63                     \\
\textbf{Obnoxious}     & \textbf{0.64}            & \textbf{0.64}            & 0.61                     \\
\textbf{OK}            & \textbf{0.73}            & 0.70                     & 0.61                     \\
\textbf{Persuasive}    & 0.83                     & \textbf{0.84}            & 0.78                     \\
\textbf{Unconvincing}  & \textbf{0.70}            & \textbf{0.70}            & 0.61                     \\ 
\midrule
\textbf{Average}                & 0.76 & 0.77 & 0.71                     \\
\bottomrule
\end{tabular}
\caption{Recalls for various rating categories. The reason we choose recall is for making comparison with the results reported by \citet{weninger2013words}.}
\label{tab:ratingwise_metric}
\end{table}
Table~\ref{tab:ratingwise_metric} provides a clearer picture how the dependency tree based neural network performs better than the word sequence neural network. The former achieves a higher recall for most of the rating categories ($9$ out of $14$). Only in three cases (\emph{Funny}, \emph{Longwinded}, and \emph{OK}) the word sequence model achieved higher performance than the dependency tree model. Both these models performed equally well for the \emph{Obnoxious} and \emph{Unconvincing} rating category. It is important to realize that the dependency trees we extracted were not manually annotated. They were extracted using SyntaxNet which itself introduces some error. Andor et al.~\shortcite{Andor2016} described their model accuracy to be approximately $0.95$. We expected to notice an impact of this error in the results. However, the results show that the additional information (Parts of Speech tags and the dependency structure) benefited the prediction performance despite the error in annotating the dependency trees. We think the hierarchical tree structure resolves many ambiguities in the sentence semantics which is not available to the word sequence model.

We also compare our results with \citet{weninger2013words}. However, this comparison is just an approximation because the number of TED talks are different in our experiment than in \citet{weninger2013words}. The results show that the neural network models perform better for almost every rating category except \emph{Fascinating} and \emph{Obnoxious}.

A neural network is a universal function approximator~\cite{cybenko1989,hornik1991} and thus expected to perform better. Yet we think another reason for its excel is its ability to process a faithful representation of the transcripts. In the baseline methods, the transcripts are provided as words without any order. In the neural counterparts, however, it is possible to maintain a more natural representation of the words---either the sequence, or the syntactic relationship among them through a dependency tree. In addition, neural networks intrinsically capture the correlations among the rating categories. The baseline methods, on the other hand, considers each category as a separate classification problem. These are a few reasons why neural networks are a better choice for the TED talk prediction task.

\section{Conclusion}
In summary, we presented neural network architectures to predict the TED talk ratings from the speech transcripts and prosody. We provide domain specific information such as psycho-linguistic language features, prosody and narrative trajectory features to the baseline classifiers. The neural networks, on the other hand, were designed to consume mostly the raw data with a few high-level assumptions on human cognition. The neural network architectures provide state of the art prediction performance, outperforming the competitive baseline method in the literature. The average AUC of the networks are $0.83$ compared to the baseline method's AUC of $0.78$. The results also show that dependency tree based networks perform better in predicting the TED talk ratings. Furthermore, inclusion of prosody does not help as much as we expect it to be. The exact reason why this happens, however, remains to be explored in the future.

The dataset and the complete source code of this work will be freely available to the scientific community for further evaluation.\footnote{link blinded due to author anonymity}

\bibliography{tanveer_ted_2019}
\bibliographystyle{acl_natbib}

\end{document}